\documentclass{emulateapj}
\usepackage{apjfonts}
\usepackage{url}
\usepackage{multirow}
\usepackage{enumerate}
\usepackage{hyperref}
\usepackage{amsmath}
\usepackage{graphicx}

\usepackage{comment}
 \usepackage{color}           
 \definecolor{DarkGreen}{rgb}{0.0,0.45,0.0}  

\newcommand{\sat}[1]{\it\uppercase{#1}\rm}
\newcommand{\fig}[1]{Figure~\ref{#1}}

\newcommand{\rsun}[1]{${#1}\,R_\odot$}

\DeclareMathOperator{\sign}{sign}
\DeclareMathOperator{\slog}{slog}
\begin{document}

\shorttitle{Long-Duration Confined Flare} %

\shortauthors{Liu et al.}

\title{An Unorthodox X-Class Long-Duration Confined Flare}
\author{Rui Liu\altaffilmark{1}, Viacheslav S. Titov\altaffilmark{2}, Tingyu Gou\altaffilmark{1,3}, Yuming Wang\altaffilmark{1}, Kai Liu\altaffilmark{1,3}, Haimin Wang\altaffilmark{4}}

\altaffiltext{1}{CAS Key Laboratory of Geospace Environment, Department of Geophysics and Planetary Sciences, University of Science and Technology of China, Hefei 230026, China; rliu@ustc.edu.cn}
\altaffiltext{2}{Predictive Science Inc., 9990 Mesa Rim Road, Suite 170, San Diego, CA 92121, USA}
\altaffiltext{3}{Mengcheng National Geophysical Observatory, School of Earth and Space Sciences, University of Science and Technology of China, Hefei 230026, China}
\altaffiltext{4}{Space Weather Research Laboratory, Center for Solar-Terrestrial Research, NJIT, Newark, NJ 07102, USA}

\begin{abstract}
We report the observation of an X-class long-duration flare which is clearly confined. It appears as a compact-loop flare in the traditional EUV passbands (171 and 195~{\AA}), but in the passbands sensitive to flare plasmas (94 and 131~{\AA}), it exhibits a cusp-shaped structure above an arcade of loops like other long-duration events. Inspecting images in a running difference approach, we find that the seemingly diffuse, quasi-static cusp-shaped structure consists of multiple nested loops that repeatedly rise upward and disappear approaching the cusp edge. Over the gradual phase, we detect numerous episodes of loop rising, each lasting minutes. A differential emission measure analysis reveals that the temperature is highest at the top of the arcade and becomes cooler at higher altitudes within the cusp-shaped structure, contrary to typical long-duration flares. With a nonlinear force-free model, our analysis shows that the event mainly involves two adjacent sheared arcades separated by a T-type hyperbolic flux tube (HFT). One of the arcades harbors a magnetic flux rope, which is identified with a filament that survives the flare owing to the strong confining field. We conclude that a new emergence of magnetic flux in the other arcade triggers the flare, while the preexisting HFT and flux rope dictate the structure and dynamics of the flare loops and ribbons during the long-lasting decay phase, and that a quasi-separatrix layer high above the HFT could account for the cusp-shaped structure.
\end{abstract}
 

\section{Introduction}
Solar flares are explosive manifestation of energy release in the solar atmosphere. Some have a strong link to coronal mass ejections (CMEs). Although such \emph{eruptive flares} are among the most important space weather-relevant events on the Sun, there are as many \emph{confined flares}, e.g.,  \citet{andrews03} reported that approximately 40\% of M-class flares between 1996 and 1999 are not associated with CMEs. In some rare occasions, even the most energetic, X-class flares (up to X3 in the literature) proceed without CMEs \citep[e.g.,][]{gaiz98, wz07}. Hence, we cannot ignore confined flares, which span a wide energy range and represent a large population, if we are to understand the physical mechanism of flares and their relationship with CMEs. 

From an observational point of view, confined flares are often characterized by an impulsive light curve in soft X-rays (SXRs), indicating that cooling dominates after the initial impulsive energy release, whereas eruptive flares have a gradual decay phase which lasts for hours, also known as long-duration-event (LDE) flares \citep{sheeley83, wh87}, suggesting that energy release continues after the impulsive phase. A further morphological distinction is that confined flares usually exhibit a simple, compact loop in SXRs, and do not have a cusp-shaped structure as often seen in LDE flares \citep{sm11}, in which the temperature is higher near the edge of the cusp-shaped structure \citep{tsuneta92, tsuneta96, tsuneta97}. However, some impulsive flares can be eruptive \citep[e.g.,][]{nh01}, and in a rare case when impulsive and LDE flares are `lumped' together, it is impossible to pinpoint with certainty which flare is responsible for the CME \citep{goff07}. 

Studies of eruptive flares have converged to a standard model, which evolved from the concepts of \citet{carmichael64}, \citet{sturrock66}, \citet{hirayama74}, and \citet{kp76}. In this model, a rising flux rope above the polarity inversion line (PIL) stretches the overlying field lines, resulting in the formation of a current sheet underneath, where magnetic reconnection heats the local coronal plasma and accelerates particles. These two processes produce thermal conduction fronts and precipitating nonthermal particles to heat the chromospheric footpoints of the newly reconnected field lines. Owing to this impulsive heating, chromospheric plasma evaporates (or ablates) and fills the reconnected flux tubes with over-dense heated plasma, which forms SXR flare loops in excess of 10 MK. Once the flare loops cool down via thermal conduction and radiation, they become visible successively in cooler EUV passbands and eventually in H$\alpha$. \citet{kp76} further predicted a continuous rise of the reconnection site, due to the rising flux rope. Consequently, the newly reconnected field lines beneath the reconnection site have an increasingly larger height and wider footpoint separation. More recent observations confirm the overall picture of the standard model but also demonstrate its insufficiency \citep[see the review by][]{benz08}. 

This standard model was further elaborated by \citet{tsuneta96, tsuneta97} and \citet{shibata95} based on \textit{Yohkoh} observations. In \citet{tsuneta97}, a distinct X-shaped structure is observed beneath a rising plasmoid and above a bright SXR flare loop. Inverse V-shaped hot ridges (15--20 MK) are located above the SXR loop and below the X-point, supposedly heated by a pair of slow shocks. A compact hot source at the SXR loop top is seen in hard X-rays (HXRs), and its counterpart associated with the plasmoid is seen in SXRs (15 MK). Both sources are assumed to be heated by fast (perpendicular) shocks, which form as the downward and upward reconnection outflows collide with the SXR loop and the plasmoid, respectively. This depicts a `complete' X-type reconnection geometry \citep{petschek64}. In particular, the HXR loop-top source is observed in several impulsive compact-loop flares \citep{masuda94, masuda95}, in which SXR plasmoid ejections are also detected high above the SXR loop \citep{shibata95}. This suggests that the main energy release takes place not inside the compact flare loop, as previously thought \citep[e.g.,][]{us88}, but above it, like the LDE flares. Hence, it has been argued that both eruptive and confined flares can be explained by fast reconnection induced by plasmoid ejection \citep{shibata95, shibata99, sm11}. 

Several mechanisms have been proposed to trigger the eruption, including tether
cutting \citep{moore01}, breakout \citep{adk99}, flux emergence \citep{cs00}, and ideal MHD instabilities \citep[e.g.,][]{fi91,tk05,tk07}. Magnetic reconnection at current sheets forming at magnetic separatrices plays a crucial role in almost all mechanisms except ideal instabilities. On the other hand, it has been demonstrated numerically that current sheets also prone to develop at quasi-separatrix layers \citep[QSLs; e.g.,][]{aulanier05}, where field line linkage displays a rapid change but is not necessarily discontinuous as in separatrices. Observationally, flare brightenings in chromosphere are indeed closely associated with the footprint of QSLs \citep[e.g.,][]{demoulin97}, suggesting that QSLs are important locations for the buildup and release of free magnetic energy in corona.


The detailed configuration of the coronal field significantly influences how the flare process proceeds. A confined flare ensues when reconnection occurs between two groups of loops \citep[e.g.,][]{hanaoka97, nishio97, melrose97, aschwanden99, su13} or at a coronal null point with a single spine that emerges away from the fan surface anchored in a remote region \citep[e.g.,][]{masson09}. The strength of the overlying field may also play an important role in regulating the behavior of solar eruptions. It is found that a toroidal flux ring is unstable to lateral expansion if the external poloidal field $B_{\mathrm{ex}}$ decreases rapidly with height such that the decay index $n=-d\ln B_{\mathrm{ex}}/d\ln h$ exceeds 3/2 \citep{bateman78, kt06, tk07, oz10}. Confined flares associated with the failed eruption of a flux rope could be attributed to this effect \citep[e.g.,][]{tk05, alg06, wz07, guo10}.  

In this paper, we present an LDE flare that is confined. This flare also exhibits a cusp-shaped structure above a post-flare loop (PFL) system, but the dynamical processes and the temperature structure within the cusp-shaped structure are opposite to what the standard flare model predicts. In the sections that follows, we analyze the observations of the flare (Section~\ref{sec-flare}), investigate the relevant magnetic configuration (Section~\ref{sec-mag}), and offer a possible interpretation for this unorthodox flare (Section~\ref{sec-conclusion}). 

\subsection{Validation of the Confinedness} \label{ss-confine}
\begin{figure*}
\includegraphics[width=\textwidth]{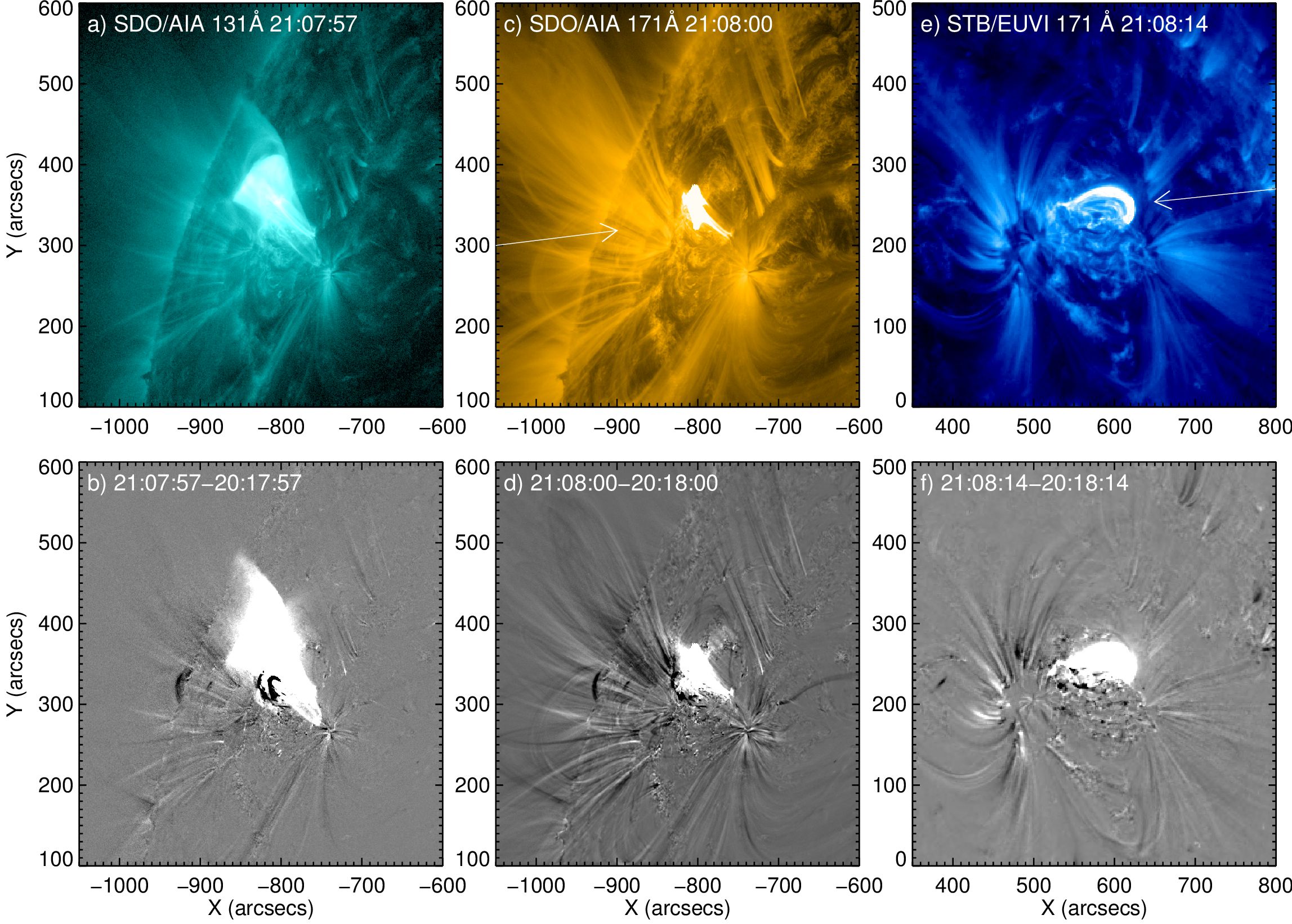} \caption{Post-flare loop system of the X1.9 flare as seen by \sat{SDO} (a--d) and \sat{stb} (e--f). In Panels (c) and (e) the arrows indicate \sat{stb}'s and \sat{sdo}'s line of sight, respectively. Bottom panels show difference images, which are obtained by subtracting `base' images taken at 20:18 UT from the corresponding images in the top panels. \label{stereo}}
\end{figure*}

\section{Observations of the Flare} \label{sec-flare}
The \sat{goes}\footnote{Geostationary Operational Environmental Satellite}-class X1.9 flare occurs at N22E63 in the NOAA\footnote{National Oceanic and Atmospheric Administration} active region 11339 on 2011 November 3. According to the \sat{goes} 1--8~{\AA} light curve, the flare starts at 20:16 UT, peaks at 20:27 UT and gradually decays to the pre-flare level at about 22:00 UT. Normally such an energetic, long-duration event is associated with a CME, but this one fails to produce a successful eruption. The flaring process is well recorded by the Atmospheric Imaging Assembly \citep[AIA;][]{lemen12} onboard the Solar Dynamic Observatory \citep[SDO;][]{pesnell12}. Among the six EUV channels of AIA,  we concentrate on the 131~{\AA} passband, which mainly contains hot \ion{Fe}{20} ($\log T=7.0$) and \ion{Fe}{23} ($\log T=7.2$) lines as well as a cool \ion{Fe}{8} ($\log T=5.6$) component. Similar features and processes are also observed in the 94~{\AA} passband, which is dominated by hot \ion{Fe}{18} lines ($\log T=6.8$) despite a cool \ion{Fe}{10} ($\log T=6.1$) component, but its signal-to-noise ratio is inferior to 131~{\AA}. In the following subsections, we verify that the flare is confined (\S\ref{ss-confine}), explore the structure and dynamic evolution of its PFL system (\S\ref{ss-loop}).

The confinedness of this LDE flare is verified by multi-satellite observations without ambiguity. \fig{stereo} shows the PFL system observed from \sat{sdo}'s viewpoint in 131 and 171~{\AA} and from the viewpoint of the ``Behind'' satellite (hereafter \sat{STB}) of  \sat{stereo}\footnote{Solar Terrestrial Relations Observatory} in 171~{\AA}. It is obvious that the PFL is seen in a ``face-on'' view by \sat{stb} and ``edge-on'' by \sat{sdo}. In the cold passbands (e.g., 171~{\AA}), a compact loop ($\sim 50$ Mm high in projection) is observed as expected from a confined flare; in the hot passbands (131 and 94~{\AA}), however, one can see a higher arch with an overlying cusp-shaped structure, which is a distinct feature of LDE flares \citep{sm11}. The arch is about 100 Mm high in projection, spanning about 60 Mm on top. Initially the cusp point is about 150 Mm high in projection and later the height increases up to about 200 Mm. In the corresponding difference images (bottom panels of \fig{stereo}), no obvious coronal dimming can be seen in the neighborhood of the flaring region.

\begin{figure*}
\includegraphics[width=\textwidth]{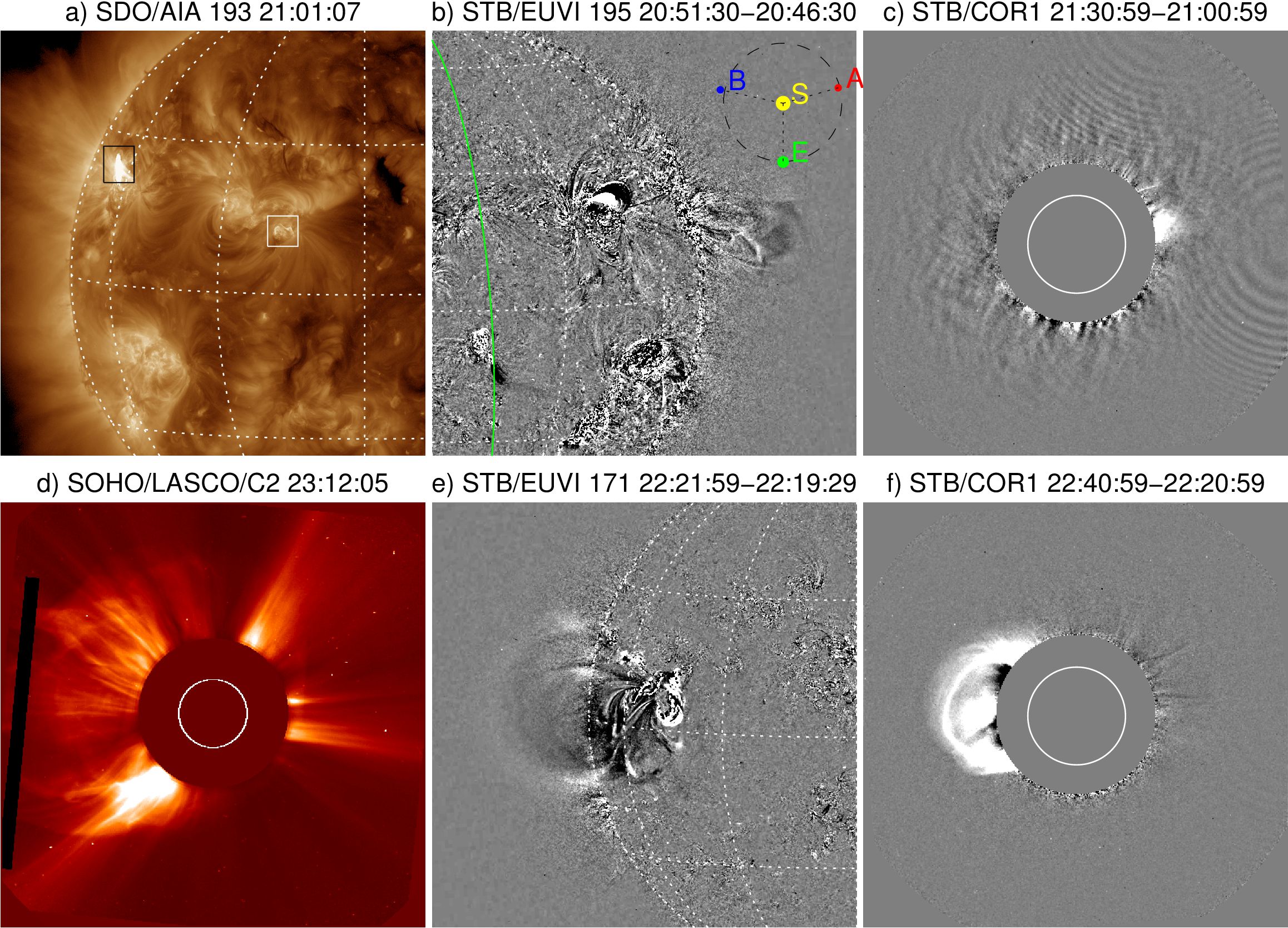} \caption{Two coronal disturbances detected within three hours after the onset of the X1.9 flare. Top panels show a CME-like disturbance observed by SDO (a) and STB (b and c) at around 21:00 UT, whereas bottom panels a CME observed by SOHO (d) and STB (e and f) at around 22:00 UT. In Panel (a), the black rectangle encloses the flare region and the white rectangle marks a flare-like brightening. In Panel (b), the green curve denotes the solar limb as seen by \sat{sdo}. The inset plots the positions of STA (red) and STB (blue) relative to the Sun (yellow) and Earth (green) in the plane of the Earth's orbit (dashed circle) at 22:00 UT on 2011 November 3. The dotted lines show the angular displacements from the Sun. \label{validate}}
\end{figure*}

\begin{figure*}
\includegraphics[width=\textwidth]{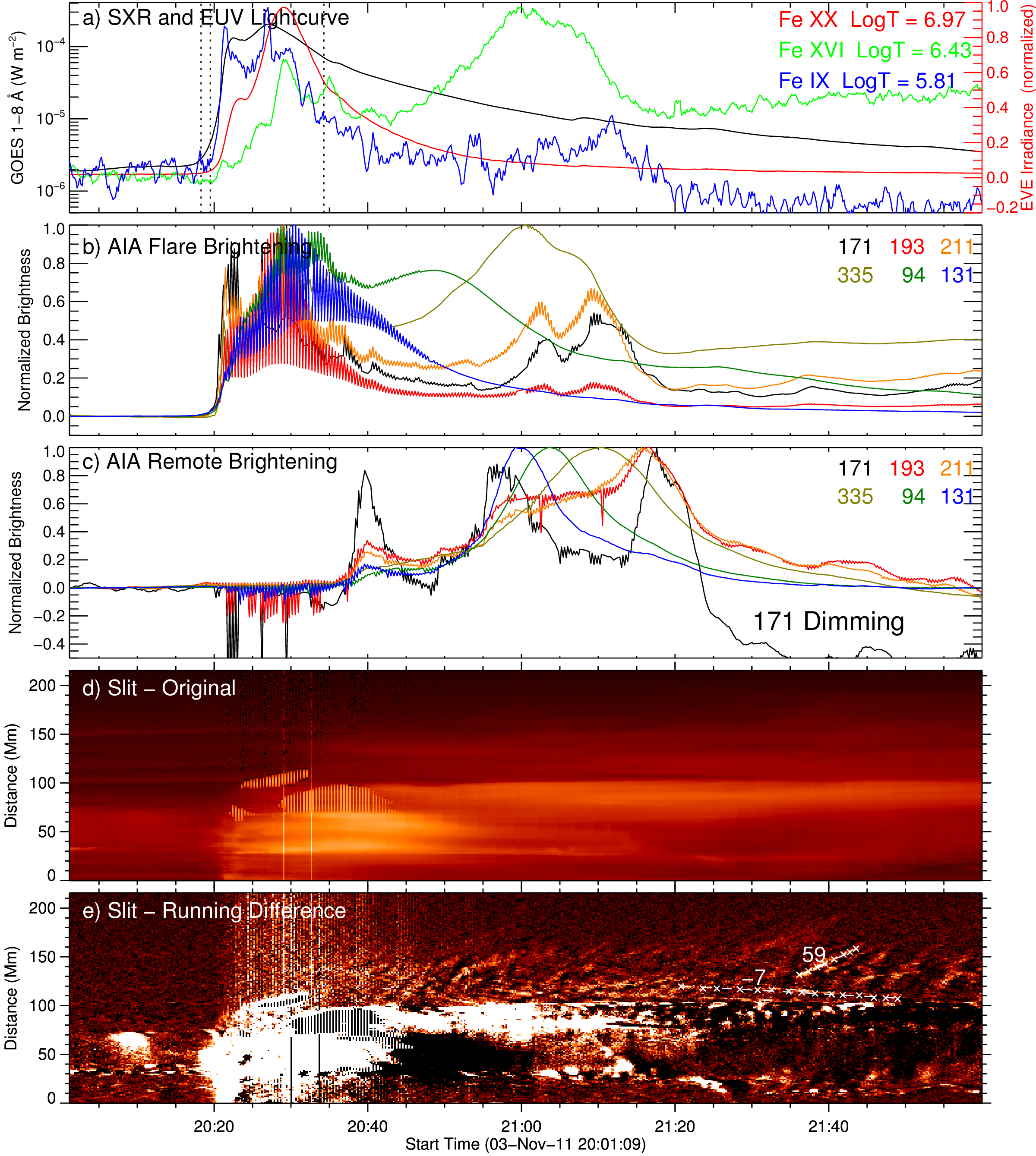} \caption{Temporal evolution of the flare. (a) \sat{goes} 1--8~{\AA} flux (black; scaled by the left y-axis) and EVE irradiances at 131~{\AA} (\ion{Fe}{20}; red), 335~{\AA} (\ion{Fe}{16}; green) and 171~{\AA} (\ion{Fe}{9}; blue), which are normalized and scaled by the right y-axis. Dotted lines mark the time instants when the three 1600~{\AA} images in Fig.~\ref{fig-uv}(a--c) are taken. (b) and (c) show the mean brightness (normalized) within the black and white boxes in \fig{validate}(a), respectively, at six EUV wavelengths. After about 21:25 UT, a dimming in 171~{\AA} below the pre-brightening background (zero) can be seen in (c). (d) and (e) show the evolution seen through the slit in \fig{loop}(d), with the former displayer in a logarithm scale and the latter made in a running difference approach. Linear fitting speeds along two representative strips (`x' symbols) are denoted in km~s$^{-1}$. \label{slit}}
\end{figure*}

We further examine the white-light images taken by coronagraphs onboard \sat{soho}\footnote{Solar and Heliospheric Observatory} (12-min cadence) and the two \sat{stereo} satellites (5-min cadence), ``Ahead'' (hereafter \sat{STA}) and ``Behind'' (\sat{STB}). The satellite positions are plotted in the inset of \fig{validate}(b). Note that both \sat{sdo} and \sat{soho} are in orbit near the Earth. Since AR 11339 is located away from the disk center from both \sat{soho}'s and \sat{stb}'s viewpoint, the chance that a CME originating from it is missed in coronagraph observations is minimal, despite a data gap in \sat{soho}'s Large Angle and Spectrometric Coronagraph (LASCO/C2; 2--\rsun{6}) during 22:00--23:12 UT. Within the 3-hr time window after the onset of the X1.9 flare, we find only one CME and a CME-like disturbance, neither of which originates from the target active region. The CME can be seen by all three satellites (see \sat{soho} and \sat{stb} observations in the bottom panels of \fig{validate}). Its source region is located behind the limb from \sat{soho}'s viewpoint and close to the east limb from \sat{stb}'s viewpoint (\fig{validate}(e)). The CME-like disturbance is only briefly captured by the COR1 coronagraph onboard \sat{stb} from 21:06 till 22:06 UT (\fig{validate}(c)). It originates from an expanding loop system located in AR 11336 (N12E12), to the west of the flaring region. From \sat{sdo}'s viewpoint, the expanding loops are apparently anchored at a flare-like brightening as enclosed by a white box in \fig{validate}(a); from \sat{stb}'s viewpoint, the expanding loops can be seen in 195~{\AA} above the west limb (\fig{validate}(b)). The mean brightness within the white box peaks successively in AIA 131~{\AA} (10 MK), 94~{\AA} (6.3 MK), 335~{\AA} (2.5 MK), 211~{\AA} (2.0 MK) and 193~{\AA} (1.6 MK) at 20:59:10, 21:03:50, 21:10:40, 21:15:37 and 21:16:33 UT, respectively (\fig{slit}(b)), in a decrease order of the peak-response temperature of the individual channel, indicating an on-going cooling process. A strong dimming in AIA 171~{\AA} (\fig{slit}(c)) exists during the time period when the CME-like disturbance is seen by \sat{stb}/COR1, indicating that there is indeed mass ejection into higher corona. This localized dimming also corresponds to the reduced irradiance of \ion{Fe}{9} (171~{\AA}; \fig{slit}(a)) during the flare gradual phase, from the EUV Variability Experiment \citep[EVE;][]{woods12} onboard \sat{sdo}. On the other hand, the enhanced irradiance of \ion{Fe}{16} (335~{\AA}; \fig{slit}(a)), known as EUV late phase \citep{woods11, liuk13}, is dominated by the flare brightening (\fig{slit}(b)), as can be seen from the similarity between the EVE and AIA 335~{\AA} light curve, the latter of which represents the mean brightness within the black box enclosing the flare region in \fig{validate}(a). It is noteworthy that the 335~{\AA} late-phase peak is significantly higher than its counterpart during the impulsive phase, signaling that additional heating may be required.

Furthermore, we find no clear association of the X1.9 flare with any radio Type II/III bursts based on dynamic spectra obtained by the WAVES instruments (10 kHz--16 MHz) onboard both \sat{stereo} satellites and those by the ground-based Green Bank Solar Radio Burst Spectrometer (5--1100 MHz). This indicates that this flare does not involve any opening of field lines or interplanetary shocks.   

\subsection{Structure \& Dynamics} \label{ss-loop}

\begin{figure*}
\includegraphics[width=\textwidth]{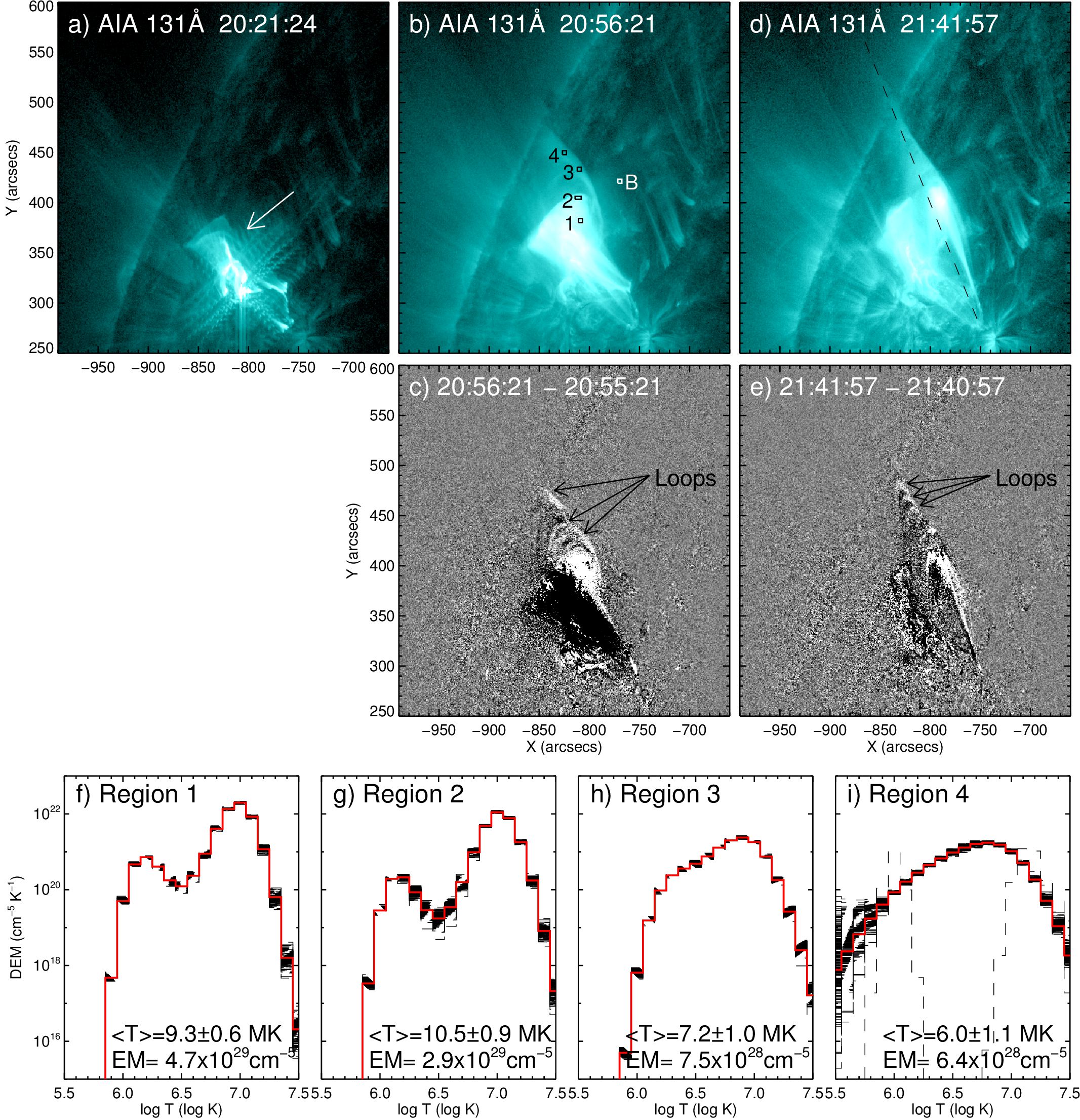} \caption{Cusp-shaped structure overlying the post-flare loop system. In Panel (a), the arrow indicates a small ejection at the onset of the flare. In Panel (b), black rectangles mark four regions where a DEM analysis is performed, and the white rectangle marks the background region for the DEM analysis. The resultant DEMs are shown in the bottom panels ((f)--(i); see the text for details). In Panel (d), the dashed line indicates the slit from which the stack plots in \fig{slit}(d) and (e) are made. Panels (c) and (e) show the difference images corresponding to Panels (b) and (d) subtracted by images taken a minute earlier, respectively. Nested loops within the cusp-shaped structure are visible in difference images. An animation of AIA 131~{\AA} images is available online. \label{loop}}
\end{figure*}

At the onset of the flare, one can see a small surge-like ejection (marked by an arrow in \fig{loop}(a)), apparently confined by the overlying loops. \citet{kc13} made a connection of this ejection, which is composed of multiple plasmoids, with drifting pulsating structures detected at 500-1200 MHz during 20:21:24--20:22:36 UT. The PFL formed after the flare peak is quite stable: its height does not increase with time, as can be seen from the stack plots (\fig{slit}(d) and (e)) made from a slit (\fig{loop}(d)) cutting across it. The cusp-shaped structure overlying the PFL appears to be diffuse and absent of fine structures. However, the running difference images (\fig{loop}(c) and (e)) reveal that it consists of multiple nested loops which are undergoing upward as well as downward motions all the time (see the animation accompanying \fig{loop}). 

One can see that the cusp-shaped structure initially give a slightly tilted ``face-on'' view (\fig{loop}(b) and (c)), but later on it tilts further toward the line of sight to appear almost ``edge-on''(\fig{loop}(d) and (e)) as the PFL. Meanwhile, the cusp point gradually increases in height and slowly moves westward. We place a virtual slit intersecting the cusp-shaped structure and perpendicular to the top of the PFL, and make it slide slowly westward to pass the cusp point all the time (see the animation accompanying \fig{loop}). The resultant stack plot is shown in a logarithm scale in \fig{slit}(d), and that obtained in a running difference approach in \fig{slit}(e). In our case, each image is subtracted by the image acquired at 60 s earlier, which gives a satisfying contrast. We have tried various time differences, from 12 to 84 s, which yield similar results but different contrast. Unlike the recent AIA observations of loop contraction below the cusp point \citep{lcp13, liu13mnras}, multiple loops above the PFL rise toward the cusp point at tens of km~s$^{-1}$ and disappear approaching the cusp point, which results in numerous positively-sloped strips above the PFL in the stack plot (\fig{slit}(e)), typically lasting several minutes. A few loops shrinking toward the PFL at a few kilometers per second constitute the negatively-sloped strips in the stack plot, which last tens of minutes, presumably representing a relaxation process. The rising proceeds in a more or less meandering fashion, such that the positively-sloped strips look diffuse and not exactly straight. The fact that numerous episodes of loop rising are detected over the whole gradual phase corroborates the irrelevance of the remote brightening detected in AR 11336, which is impulsive and dominated by cooling (\fig{slit}(b)). 

\begin{figure*}
\includegraphics[width=\textwidth]{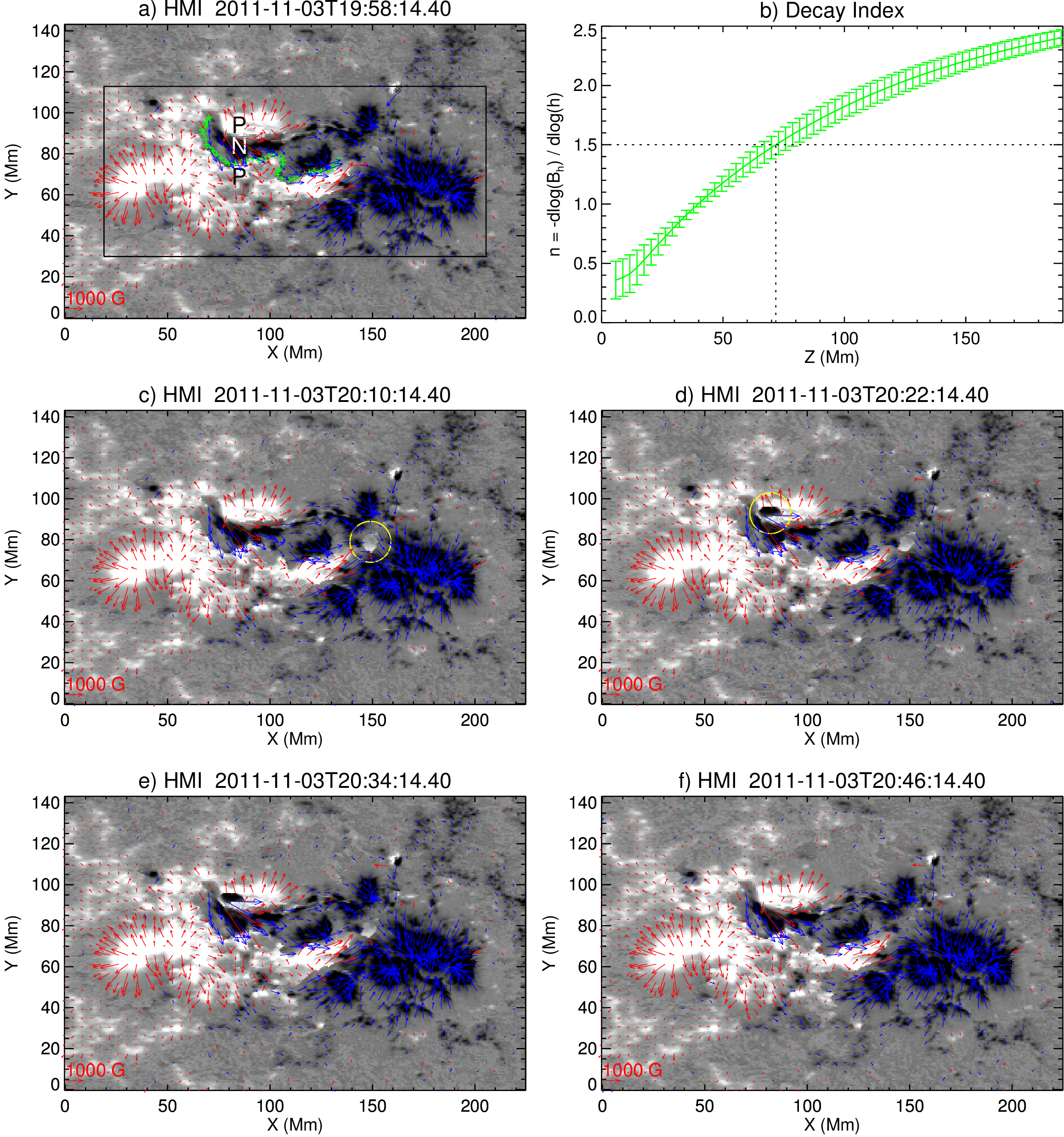} \caption{Sequence of HMI vector magnetograms across the time period of the flare. Red (blue) arrows indicate the horizontal component of the photospheric field originating from positive (negative) polarities of the vertical component. The rectangle in (a) denotes the region in which the $Q$-factor is calculated. Panel (b) shows the decay index $n=-d\log(B_h)d\log(h)$ with height, which is averaged over the hand-picked points along the PIL of interest (`+' symbols in (a)). The yellow circles in (c) and (d) mark the location of new flux emergence.  \label{fig-hmi-sequence}}
\end{figure*}

We use \texttt{xrt\_dem\_iterative2} in SolarSoft to compute the differential emission measure (DEM) in four selected regions (black rectangles in \fig{loop}(b)). This code was originally designed for Hinode/X-ray Telescope data \citep{golub04, weber04}, and recently modified to accommodate the AIA temperature response \citep[e.g.,][]{winebarger11,cheng12}. We select a nearby quiet-Sun region as the background (white rectangle labeled `B' in \fig{loop}(b)), which is free of coronal structures in any passband. The best-fit DEM solutions to the mean fluxes acquired by six AIA EUV channels are shown in the bottom panels of \fig{loop} (red histogram-style solid curves). To give a sense of uncertainty, 250 Monte Carlo (MC) simulations (black dashed curves) are performed for each best-fit DEM, by randomly varying the observed fluxes up to the uncertainties estimated by \texttt{aia\_bp\_estimate\_error} in SolarSoft. An average temperature is given, weighted by the best-fit DEM \citep{cheng12}. Its uncertainty is approximated with the standard deviation of the DEM-weighted temperatures of the 250 MC simulations. It can be seen that the top of the PFL has the highest temperature ($\sim 10$ MK; \fig{loop}(g)), while the cusp-shaped structure overlying the PFL is significantly cooler (\fig{loop}(h) and (i)). Furthermore, it is cooler at higher altitudes above the PFL. Thus, this cusp-shaped structure must not be the locus of main energy release as in the standard model.

\section{Magnetic Configuration} \label{sec-mag}
\begin{figure*}
\includegraphics[width=\textwidth]{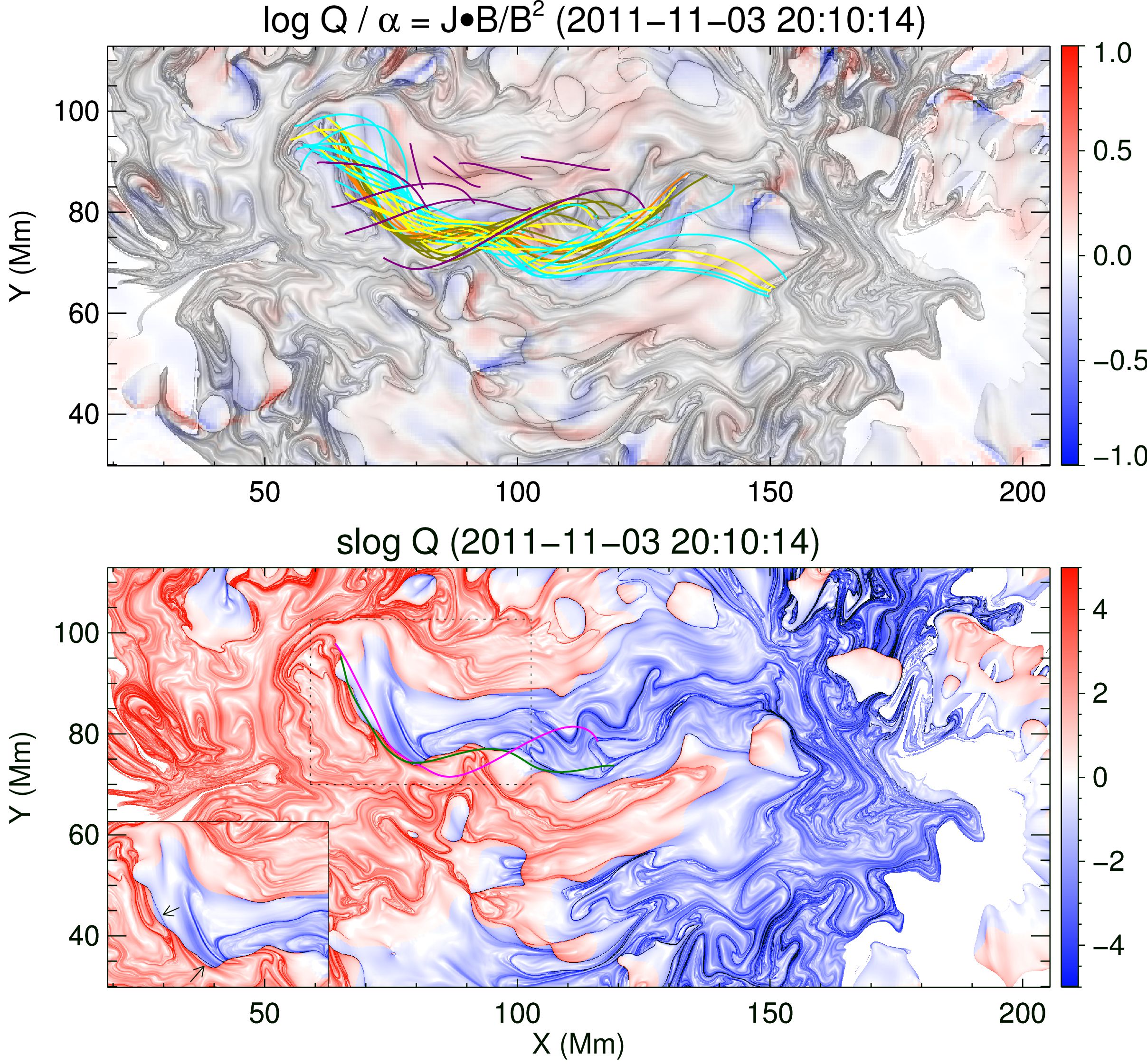} \caption{Magnetic structures as revealed by the Q-map. Top panel: a pre-flare $\log Q$-map (black and white) blended with the corresponding $\alpha$-map (red and blue). BPSS field lines tangent to the three BP subsegments are in three different colors, orange, olive and cyan, respectively; representative field lines of the flux rope are shown in yellow; purple lines indicate two sheared arcades. Bottom panel: $\slog Q$ map superimposed with the BPSS separators (green and magneta lines). The rectangular region as indicated by dotted lines is reproduced in the bottom-left corner, in which two BP separations are marked with arrows.  \label{fig-hmi-init}}
\end{figure*}

To understand the actual scenario of this event, we investigate its magnetic configuration by utilizing a sequence of HMI vector magnetograms (Fig.~\ref{fig-hmi-sequence}) obtained from 19:58 UT till 20:46 UT on 2011 November 3 at 12-min cadence, which span a time period across the flare. These disambiguated vector magnetograms have been remapped using a cylindrical equal area projection (CEA), sampling at 0.03 deg, and presented as $(B_r,B_\theta, B_\phi)$ in a heliocentric spherical coordinate, corresponding to $(B_z,-B_y,B_x)$ in the heliographic coordinates \citep{sun13}. The vector magnetograms are ``pre-processed'' to best suit the force-free condition \citep{wiegelmann06} before being taken as the photospheric boundary to extrapolate a nonlinear force-free field (NLFFF) using the ``weighted optimization'' method \citep{wiegelmann04}. This code has been optimized recently for HMI data by taking into account measurement errors in photospheric field \citep{wiegelmann12}. Our calculation is performed using 2 by 2 rebinned magnetograms within a box of $420\times 320\times 320$ uniform grid points, whose FOV is slightly larger than what is shown in Fig.~\ref{fig-hmi-sequence}. Using a $10\times10$ smaller cell size, we refine the photospheric computational grid and map it along magnetic field lines to counterpart photospheric polarity regions. The coordinates of the mapped grid points are used then to calculate the squashing factor $Q$ of elemental magnetic flux tubes rooted at these polarities \citep[Fig.~\ref{fig-hmi-init}; ][]{titov02, titov07}.

It is found that the photospheric magnetic field in the flaring region has locally a tripolar structure in which an elephant-trunk-like area of negative polarity (labeled `N' in Fig.~\ref{fig-hmi-sequence}(a)) deeply intrudes into positive polarities (labeled `P' in Fig.~\ref{fig-hmi-sequence}(a)). The important structural features are revealed in the photospheric $\slog Q$ maps (e.g., bottom panel of Fig.~\ref{fig-hmi-init}), where $\slog Q \approx \sign(B_{z})\, \log Q$ at $Q \gg 2$ \citep{titov11}. The high-$Q$ lines designate in such maps the footprints of separatrix surfaces and quasi-separatrix layers (QSLs). In particular, the elephant-trunk-like area of negative polarity is divided in half by such a high-$Q$ line, with the respective halves belonging to two adjacent sheared arcades (purple lines in the top panel of Fig.~\ref{fig-hmi-init}). The torsion parameter $\alpha=({\mathbf{j} \cdot \mathbf{B}})/ B^2$ has opposite signs in these arcades (top panel of Fig.~\ref{fig-hmi-init}), implying that their axial currents are oppositely directed. A hyperbolic flux tube \citep[HFT;][]{titov02} that consists of two QSLs adjoining each other separates the two sheared arcades. The HFT has a T-type junction passing through the joint arcades' apex. The footprint of this junction corresponds to the above-mentioned high-$Q$ line.

Within these arcades, only the southern one contains a flux rope, which consists of multiple braided strands (yellow lines in the top panel of Fig.~\ref{fig-hmi-init}). This element of the configuration is revealed by an extended segment of the PIL that is characterized by a localized saturation of the red and blue colors (bottom panel of Fig.~\ref{fig-hmi-init}). This segment is a topological feature \citep{seehafer86} called ``bald patch" \citep[BP;][]{titov93}, where coronal field lines are tangent to the photosphere and directed from negative to positive polarity of the photospheric $B_z$ distribution. These magnetic field lines form the separatrix surface (BPSS; orange, olive and cyan lines in the top panel of Fig.~\ref{fig-hmi-init}), which wraps around the flux rope, as previously demonstrated for different models of flux-rope configurations by \citet{demoulin96}, \citet{td99}, and \citet{titov08}. Furthermore, the indicated BP bifurcates into a pair of high-$Q$ lines in two occasions by forming two husk-like shapes (marked by arrows in the bottom panel of Fig.~\ref{fig-hmi-init}). Such features imply the presence of two BP separator field lines touching the tips of the ``husks" (deep green and magenta lines in the bottom panel of Fig.~\ref{fig-hmi-init}). Each of the separators rises above the bifurcated BP gap and lies at the X-type intersection of two pieces of the BPSS, which originate at two BP subsegments adjacent to the gap. Similar structure but with a single BP gap was first described by \citet{td99} for an analytical model of the configuration with one flux rope of a circular shape. Such a complex structure of the BPSS in our case is due to upward bending of the flux rope at the BP gaps. The indicated separators are those sites where a local current concentration and subsequent tether-cutting reconnection have to occur in response to displacements of the flux rope. It is clear that the location and the number of such separators must vary as the flux rope changes its shape and height. It further comes to our notice that the average decay index (Fig.~\ref{fig-hmi-sequence}(b)), $n=-d\log(B_h)d\log(h)$\footnote{$B_h$ indicates the horizontal component of a potential field obtained with the Green's function method.}, along this flux rope only exceeds the threshold value of 1.5 above $\sim\,$70 Mm for the torus instability to function \citep{kt06}. This may account for the confinement of the flux rope, which lies below $\sim\,$25 Mm.  

\begin{figure*}
\includegraphics[width=\textwidth]{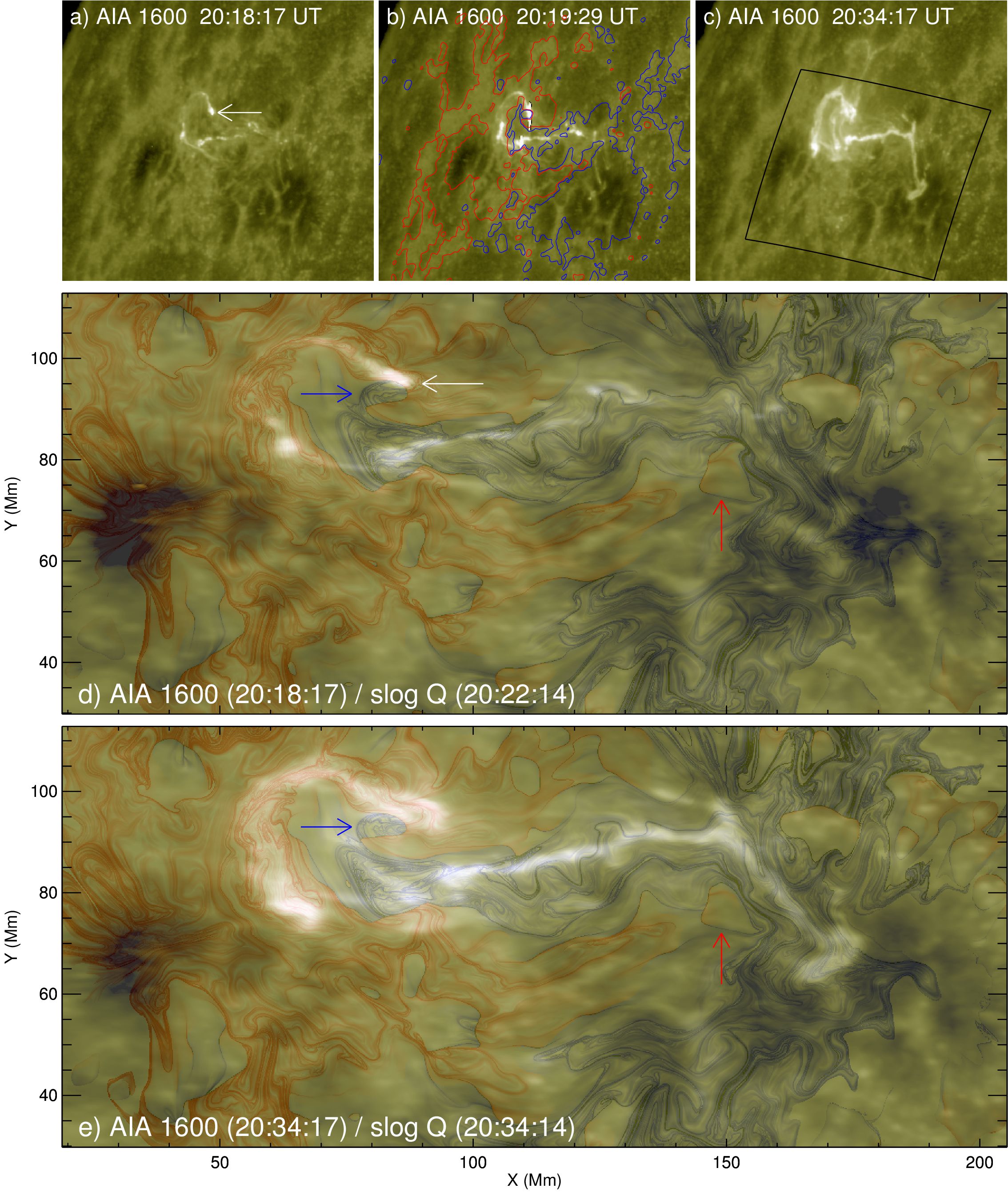} \caption{Comparison between flare ribbons and $\slog Q$ maps. (a-c) Evolution of the flare ribbons. (b) is superimposed with contours of the local $B_r$ at $\pm 200$ G, with red (blue) colors indicating positive (negative) polarities. In (c) the FOV of the $\slog Q$ maps is indicated by a warped rectangle.  (d) A blend of a UV 1600~{\AA} image taken at the onset of the flare (same as (a)) with the $\slog Q$ map calculated for the HMI vector magnetogram acquired at approximately the same time; both are remapped with the CEA projection. The most intense brightening is marked by a white arrow in (a) and (d). (e) Similar to (d) but the image (same as (c)) and $\slog Q$ map are obtained when the flare ribbons have been fully developed. The positive (negative) parasitic element is marked by a red (blue) arrow in (d) and (e). \label{fig-uv}}
\end{figure*}

\begin{figure*}
\includegraphics[width=\textwidth]{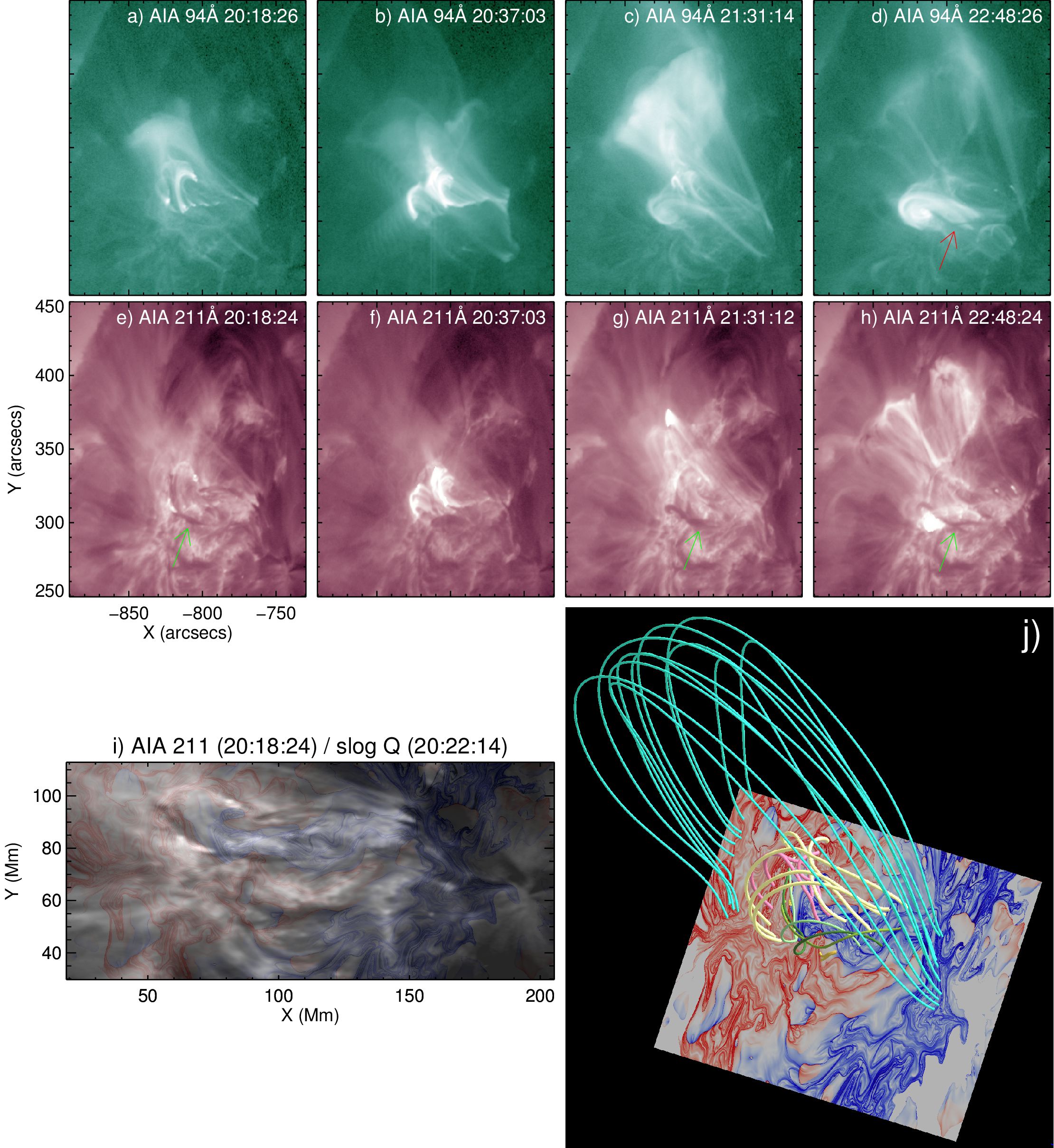} \caption{The flux rope. Panels (a--h) show a sequence of AIA 94~{\AA} (\ion{Fe}{18}; $\log T=6.8$) and 211~{\AA} (\ion{Fe}{14}; $\log T=6.3$) ) images. The dark filament associated with the flux rope is marked by green arrows in (e--h) and a red arrow in (d). In Panel (i) the 211~{\AA} image at 20:18:24 UT (same as (e)) is remapped with the CEA projection and blended with a $\slog Q$ map calculated for the HMI vector magnetogram at 20:22:14 UT. Panel (j) shows a three-dimensional view of selected field lines of the HFT (pink and yellow), the QSL (cyan) originating from the hook of the flare ribbon half circling the positive parasitic element (see Fig.~\ref{fig-uv}), and the flux rope (light green).  \label{fig-fluxrope}}
\end{figure*}

The evolution of the photospheric field towards the flare can be highlighted by the emergence of two parasitic polarity elements, as marked by yellow circles in Fig.~\ref{fig-hmi-sequence}. The negative parasitic element that appears at the onset of the flare (Fig.~\ref{fig-hmi-sequence}(d)) is closely associated with the most intense flare brightening in UV 1600~{\AA} at that time (marked by an arrow Fig.~\ref{fig-uv}(a) and (d)). Hence, we suggest that the flare is triggered by this emergence of new magnetic flux within the northern arcade. The flare brightening then quickly extends along two different paths (Fig.~\ref{fig-uv}(b)), roughly following the high-Q lines (Fig.~\ref{fig-uv}(b) and (e)). No ribbon separation movement is observed, distinct from typical eruptive flares. The flare ribbon that follows the high-Q line dividing the elephant-trunk-like area of negative polarity develops later a `hook' that half circles the positive parasitic element (Fig.~\ref{fig-uv}(c) and (e)). Field lines anchored at this hook form a QSL (cyan lines in Fig.~\ref{fig-fluxrope}(j)) high above the T-type HFT (pink and yellow lines in Fig.~\ref{fig-fluxrope}(j)).

Guided by the photospheric $\slog Q$ maps, we are able to identify the flux rope with a arc-shaped dark filament (Fig.~\ref{fig-fluxrope}(e)) aligned along the PIL that harbors the BPs (Fig.~\ref{fig-fluxrope}(i)). The filament survives the X-class flare, which is consistent with the persistent existence of the BPs. The dark filament takes on a reverse S-shape after the flare (Fig.~\ref{fig-fluxrope}(g) and (h)), enclosed by bright emissions in AIA hot filters (Fig.~\ref{fig-fluxrope}(c) and (d)). In its later evolution, the flux rope becomes brightened repeatedly with a series of confined flares in this active region, e.g., the GOES C5.8 flare peaking at 22:35 UT (Fig.~\ref{fig-fluxrope}(d)). This hot layer of plasma appears to be associated with the BPSS structures (top panel of Fig.~\ref{fig-hmi-init}). One can further see that the HFT lines (Fig.~\ref{fig-fluxrope}(j), pink and yellow) compare favorably with the flaring loops during the impulsive phase (Fig.~\ref{fig-fluxrope}(b) and (f)), and that the QSL lines (Fig.~\ref{fig-fluxrope}(j)), cyan) originating from the western `hook' of the flare ribbons (Fig.~\ref{fig-uv}(c) and (e)) share similarity with the post-flare arcade, as well as the cusp-shaped structure above the arcade, during the gradual phase (Fig.~\ref{fig-fluxrope}(b) and (c)).

It is worth mentioning that the potential field extrapolated from the photospheric $B_z$ gives a similar large-scale HFT (not shown here), suggesting that such structural skeletons are quite robust. This robustness has been demonstrated by earlier studies employing other coronal field models \citep{demoulin06,demoulin07}. However, one should keep in mind that the current analysis is based on a not-so-robust assumption that force-free conditions are applicable to the photospheric field that serves as the boundary for the field extrapolation.

\section{Conclusion} \label{sec-conclusion}
\begin{figure}
\includegraphics[width=0.45\textwidth]{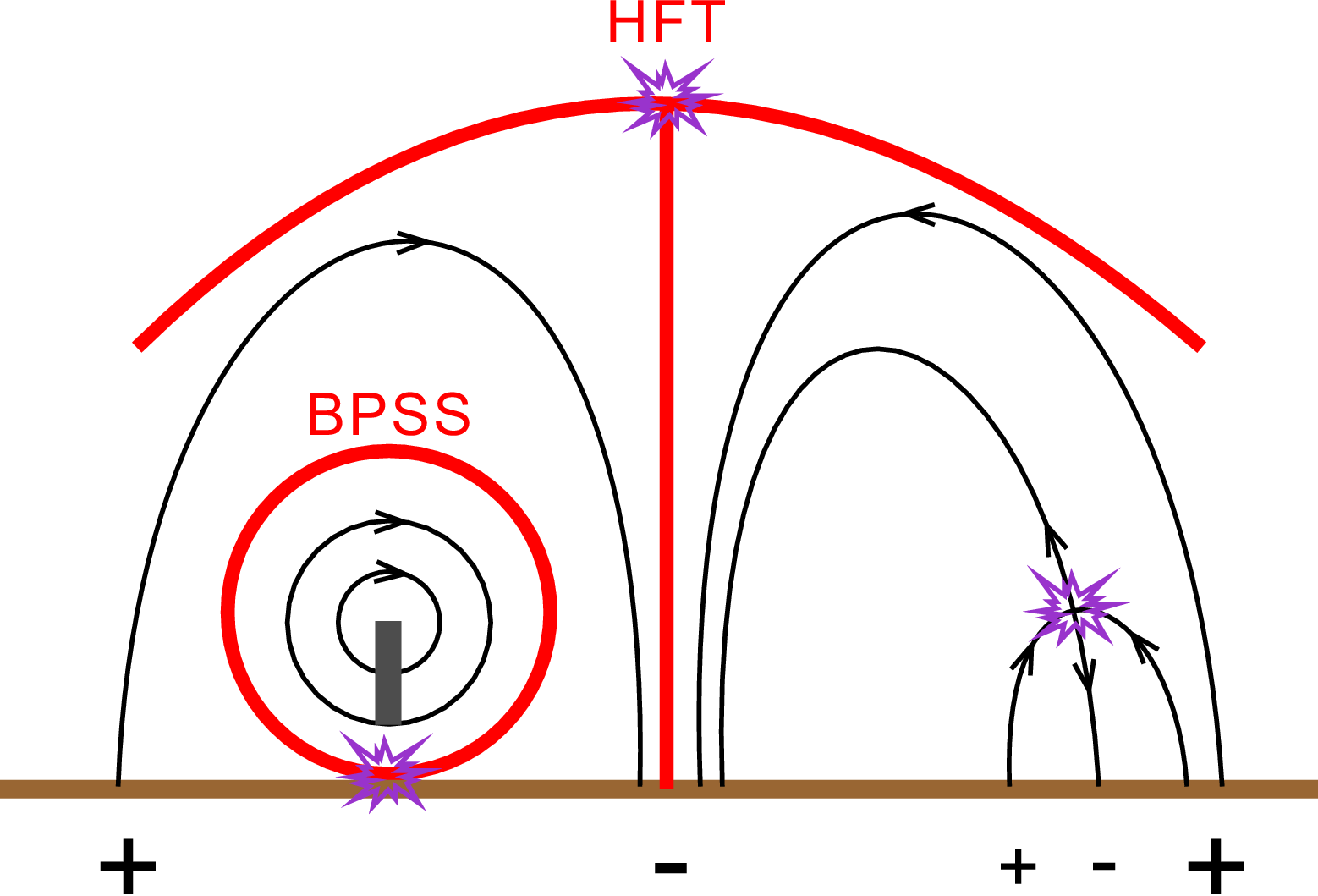} \caption{Two-dimensional schematic description of the magnetic configuration in a cross section passing though the parasitic element of negative polarity (see the text for details). The gray bar denotes the filament associated with the flux rope. Magnetic reconnection is denoted by explosion shapes at the BPSS in the left arcade, the parasitic element in the right arcade, and the T-type junction above. \label{cartoon}}
\end{figure}

What is most intriguing about this confined LDE flare is that the diffuse cusp-shaped structure above the PFL is actually composed of multiple rising loops, which disappear approaching the cusp point. Numerous episodes of loop rising are detected, spanning the whole gradual phase. The temperature is highest at the top of PFL and becomes cooler at higher altitudes. These features are contrary to a typical LDE flare. Since these rising loops do not reappear in cooler passbands later, one can safely exclude cooling processes: the typical cooling time is $\sim 10^3$~s for conduction \citep[][Eq.~(7.5.1)]{aschwanden06} and $\sim 10^5$~s for radiation \citep[][Eq.~(7.5.3)]{aschwanden06}, utilizing the loop length of $(1-2)\times10^{10}$ cm and temperature of $10^7$ K, and taking a typical coronal density of $10^9$ cm$^{-3}$. The only plausible interpretation is that the cusp-shaped structure represents a QSL, across which the connectivity of field lines undergoes sudden changes, and that the rising loops within the cusp-shaped structure is a manifestation of the on-going three-dimensional magnetic reconnection at the QSL, where the field lines slip through the plasma. This interpretation is substantiated by the analysis of the photospheric Q-maps (Section~\ref{sec-mag}).

The critical magnetic structures involved in this flare are sketched in a two-dimensional diagram (\fig{cartoon}), which features a cross section passing though the negative parasitic element. In this simplified diagram, two magnetic arcades are separated from each other by a T-type HFT. A flux rope is preexistent within the left arcade, with a filament embedded at the concave-upward portion of the rope. At the onset of the flare, the emergence of a parasitic element of negative polarity in the right arcade triggers at the HFT a fast magnetic reconnection that subsequently releases the magnetic energy and stress accumulated in the arcades, as evidenced by the UV flare ribbons residing at the high Q-lines that delineate the footprint of the HFT, with reference to Fig.~\ref{fig-uv}. Magnetic reconnection at the HFT is persistent over the flare, resulting in the observed dynamics of the nested loops within the cusp-shaped structure, which might correspond to the QSL high above the T-type HFT, with reference to Fig.~\ref{fig-fluxrope}(j). Magnetic reconnection at the BPSS also heats up plasmas surrounding the filament, with reference to Fig.~\ref{fig-fluxrope}(d). The continuing energy release at the HFT and BPSS may account for the long-lasting gradual phase of this confined flare. 

We conclude that the preexisting T-type HFT and flux rope dictate the structure and dynamics of the observed loops and ribbons in this event. The flux rope fails to escape owing to a strong confining field but succeeds in producing an LDE flare with the continuing dissipation of currents concentrated at separatrix surfaces and QSLs, which has not been considered as a significant energy source during the flare gradual phase. Hence, our analysis may shed light on the so-called EUV late phase.

\acknowledgments We thank the anonymous referee for suggestions that have improved this paper. We are grateful to the \sat{sdo}, \sat{stereo} and \sat{soho} consortium for the free access to the data. R.~Liu thanks Drs.~Chang Liu and Ju Jing for help in implementing the NLFFF code, and acknowledges the Thousand Young Talents Programme of China, NSFC 41222031 and NSF AGS-1153226. This work was also supported by NSFC 41131065 and 41121003, 973 key project 2011CB811403, CAS Key Research Program KZZD-EW-01-4, the fundamental research funds for the central universities WK2080000031. The contribution of V.S.~Titov was supported by NSF SHINE program.


\end{document}